%% file: aaai24.tex
\documentclass[letterpaper]{article} % DO NOT CHANGE THIS
\usepackage{aaai24}  % DO NOT CHANGE THIS
\usepackage{times}  % DO NOT CHANGE THIS
\usepackage{helvet}  % DO NOT CHANGE THIS
\usepackage{courier}  % DO NOT CHANGE THIS
\usepackage[hyphens]{url}  % DO NOT CHANGE THIS
\usepackage{graphicx} % DO NOT CHANGE THIS
\usepackage{natbib}  % DO NOT CHANGE THIS AND DO NOT ADD ANY OPTIONS TO IT
\usepackage{caption} % DO NOT CHANGE THIS AND DO NOT ADD ANY OPTIONS TO IT
\usepackage{algorithm}
\usepackage{algorithmic}
\usepackage{booktabs}
\usepackage{subcaption}
\usepackage{multirow}
\usepackage{algorithm}
\usepackage{algorithmic}
\usepackage{booktabs}
\usepackage{subcaption}
\usepackage{multirow}
\usepackage{color}
\input{math_commands.tex}

\usepackage{newfloat}
\usepackage{listings}
\DeclareCaptionStyle{ruled}{labelfont=normalfont,labelsep=colon,strut=off} % DO NOT CHANGE THIS
\floatstyle{ruled}
\newfloat{listing}{tb}{lst}{}
\floatname{listing}{Listing}
\title{What to Remember: Self-Adaptive Continual Learning for Audio Deepfake Detection}
\author{
    Xiaohui Zhang\textsuperscript{\rm 1, \rm 2},
    Jiangyan Yi\textsuperscript{\rm 1},
    Chenglong Wang\textsuperscript{\rm 1, \rm 4},
    Chuyuan Zhang\textsuperscript{\rm 1},
    Siding Zeng\textsuperscript{\rm 1},
    Jianhua Tao\textsuperscript{\rm 3}
}
\affiliations{
    \textsuperscript{\rm 1} State Key Laboratory of Multimodal Artificial Intelligence Systems, Institute of Automation, Chinese Academy of Sciences, Beijing, China\\
    \textsuperscript{\rm 2} School of Computer and Information Technology,
University of Beijing Jiaotong, Beijing, China\\
    \textsuperscript{\rm 3} Department of Automation, Tsinghua University, Beijing, China\\
    \textsuperscript{\rm 4} University of Science and Technology of China, Beijing, China.\\
    
}

\begin{document}

\maketitle

\begin{abstract}
% \color{red}
The rapid evolution of speech synthesis and voice conversion has raised substantial concerns due to the potential misuse of such technology, prompting a pressing need for effective audio deepfake detection mechanisms. Existing detection models have shown remarkable success in discriminating known deepfake audio, but struggle when encountering new attack types. To address this challenge, one of the emergent effective approaches is continual learning. In this paper, we propose a continual learning approach called Radian Weight Modification (RWM) for audio deepfake detection. The fundamental concept underlying RWM involves categorizing all classes into two groups: those with compact feature distributions across tasks, such as genuine audio, and those with more spread-out distributions, like various types of fake audio. These distinctions are quantified by means of the in-class cosine distance, which subsequently serves as the basis for RWM to introduce a trainable gradient modification direction for distinct data types. Experimental evaluations against mainstream continual learning methods reveal the superiority of RWM in terms of knowledge acquisition and mitigating forgetting in audio deepfake detection. Furthermore, RWM's applicability extends beyond audio deepfake detection, demonstrating its potential significance in diverse machine learning domains such as image recognition.
% \color{black}
\end{abstract}
\section{Introduction}
\label{introduction}
In recent years, the advancement of speech synthesis and voice conversion technologies has blurred the line between reality and fabrication \citep{Wang2018StyleTU, Wang2021ProsodyAV}. This has significantly amplified concerns about the potential misuse of audio deepfakes – synthesized audio that closely mimics genuine human speech, posing serious threats to social stability and public interests. Consequently, the pursuit of reliable audio deepfake detection mechanisms has garnered increasing attention across research domains. The landscape of audio deepfake detection has witnessed substantial growth, catalyzed by a series of challenges such as the ASVspoof challenge \citep{wu2015asvspoof, kinnunen2017asvspoof, todisco2019asvspoof, yamagishi2021asvspoof} and the Audio Deep Synthesis Detection (ADD) challenge \citep{yi2022add, DBLP:journals/corr/abs-2305-13774}. These competitions have underscored the crucial role of deep neural networks in achieving remarkable success in audio deepfake detection. With the advent of large-scale pre-trained models, audio deepfake detection has experienced significant breakthroughs, boasting impressive performance on publicly available datasets \citep{tak2022automatic, martin2022vicomtech, lv2022fake, wang2021investigating}.\par
\begin{figure}
\begin{center}
\begin{subfigure}[ht]{0.495\linewidth}
\begin{center}
\includegraphics[width=1.\linewidth]{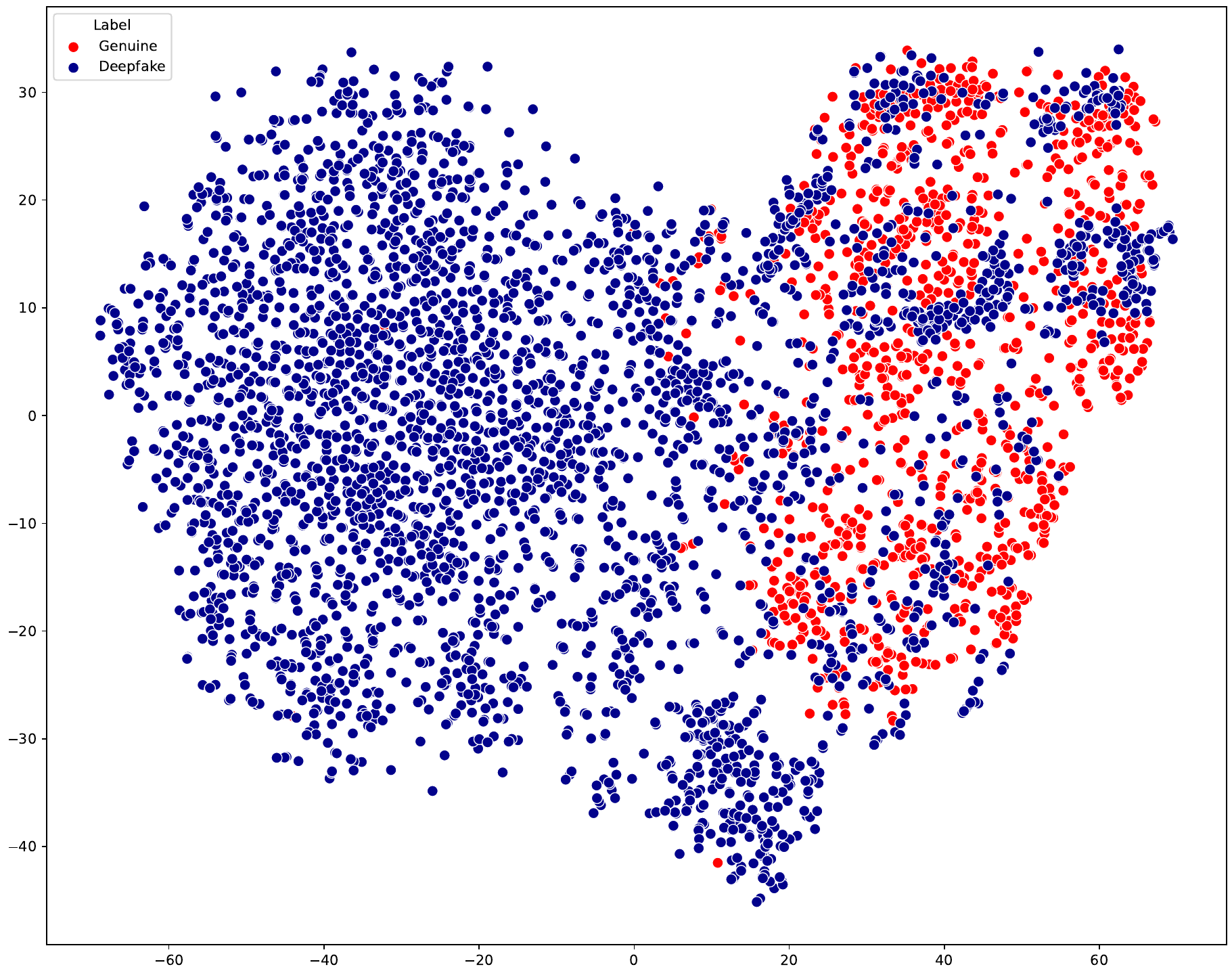}
\caption{}
\end{center}
\end{subfigure}
\begin{subfigure}[ht]{0.495\linewidth}
\begin{center}
\includegraphics[width=1.\linewidth]{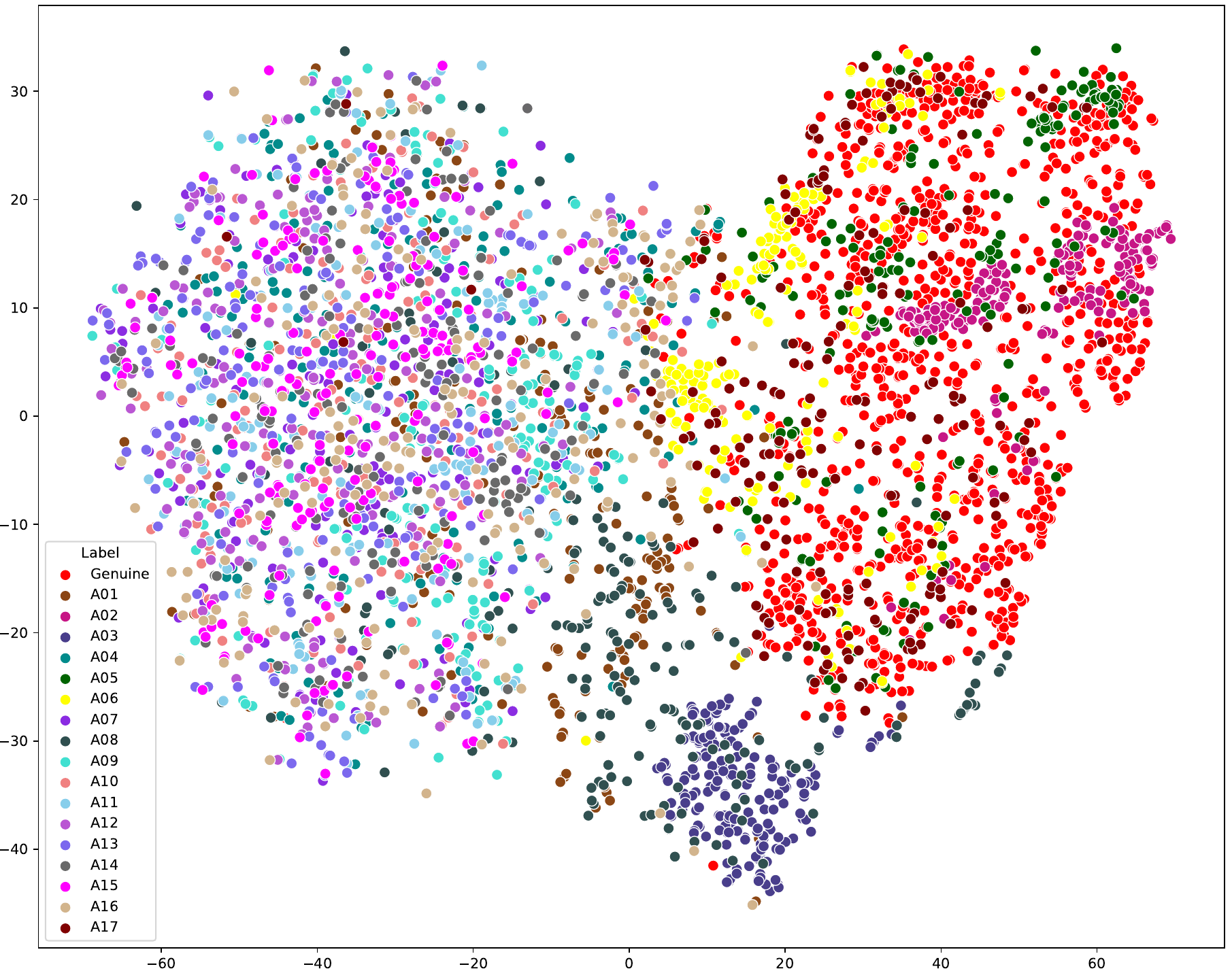} 
\caption{}
\end{center}
\end{subfigure}
\caption{The t-SNE \citep{van2008visualizing} visualization of genuine and various deepfake audio in the ASVspoof2019LA dataset visualized using Linear Frequency Cepstral Coefficients (LFCC) feature \citep{DBLP:conf/interspeech/SahidullahKH15}. All sentences are first blocked with a 20 ms Hamming window with a 10 ms shift, and then unify the frame numbers of all features into 100. (a) shows the comparison of feature distribution between genuine and deepfake audio, and (b) is the feature visualization of all audio types, including genuine and various deepfake audio.}
\label{dis_asv}
\end{center}
\end{figure}
However, existing detection models face a critical challenge, namely degraded performance when dealing with new types of deepfake audio. This challenge underscores the need for strategies to enhance the adaptability and resilience of audio deepfake detection models. To this end, two primary approaches have emerged. \citep{zhang2021empirical, zhang2021one}. The first approach involves extracting more discriminative features and developing robust model architectures to bolster the robustness of detection models against new types of deepfake audio. This strategy proves valuable in scenarios where access to data representing new types of deepfake attacks is not accessible, such as during the initial stages of encountering an unknown attack. In contrast, the second approach leverages the principles of continual learning, enabling deepfake detection models to sequentially learn from newly collected data \citep{ma2021continual}. This method capitalizes on the advantages of maintaining proficiency in detecting known deepfake types while simultaneously enhancing detection accuracy for emerging, unencountered attack.\par
For those general and widely-used continual learning algorithms, experience replay \citep{DBLP:conf/iclr/ChaudhryRRE19, DBLP:conf/eccv/PrabhuTD20} has demonstrated success across diverse domains. However, its applicability in audio deepfake detection is challenged by the acquisition of old data. Alternatively, regularization-based continual learning methods offer a more flexible approach by obviating the need for prior data. Among these methods, the Detecting Fake Without Forgetting (DFWF) \citep{ma2021continual} approach stands as the pioneering solution tailored specifically for audio deepfake detection. While DFWF exhibits notable strengths in overcoming forgetting, it still deteriorates learning performance in the context of new attack types compared to finetuning.\par
To address this limitation, we propose a continual learning approach named Radian Weight Modification (RWM) for audio deepfake detection. Most fake audio detection datasets are under clean conditions, where the genuine audio has a more similar feature distribution than the fake audio \citep{ma2021continual}, as shown in Fig. \ref{dis_asv}, and they can be seen as a whole from the same dataset or generated by the experienced-replay method on different tasks. From the view of replay, data replayed on the new task should be trained without any additional modification. Based on the above inference, it is more effective for genuine audio on new datasets to be trained with as little modification as possible. Drawing inspiration from the disparities in feature distribution between the genuine and various types of fake audio, RWM splits all classes into two groups and leverages a self-attention mechanism to enable the model to learn optimal gradient modification directions based on the current input batch. Specifically, the algorithm adapts the gradient direction based on the feature similarity between different tasks. By categorizing classes into two groups—those with compact feature distributions across tasks and those with more disparate distributions—we employ distinct strategies. When confronted with data featuring distinct characteristics across tasks, such as various types of fake audio, the algorithm guides the model to adopt a direction orthogonal to the previous data plane, ensuring preservation of learned knowledge during adaptation to new deepfake algorithms. Conversely, for data exhibiting similar features, exemplified by genuine audio, the algorithm encourages the model to learn a gradient modification direction aligned with the previous data plane, thus minimizing the interference from gradient modification. The experiments conducted on audio deepfake detection demonstrate the superiority of our proposed approach over several mainstream continual learning methods, including Elastic Weight Consolidation (EWC) \citep{kirkpatrick2017overcoming}, Learning without Forgetting (LwF) \citep{li2017learning}, Orthogonal Weight Modification (OWM) \citep{zeng2019continual} and DFWF, in terms of knowledge acquisition and mitigating forgetting. In addition, RWM can also be easily generalized to other machine learning fields. Our experiments conducted on image recognition underscore its potential significance across diverse machine learning domains. Furthermore, the utilization of the RWM method obviates the requirement for accessing previously stored data, thereby conferring a wide-ranging applicability in diverse domains of practical significance. \par
In summary, we make the following contributions.\par 
\begin{itemize}
\item
We propose a continual learning approach for audio deepfake detection that enables the model to learn discriminative information for classification on each task while autonomously optimizing the gradient direction for continuous learning across different tasks based on the similarity of feature distributions.
\end{itemize}
\begin{itemize}
\item
%\color{blue}
Although our method is inspired by the difference of feature distribution in audio deepfake detection, RWM can be applied to various machine learning fields, such as image recognition, and is not limited to any specific domain.
\end{itemize}
The code of the RWM has been uploaded in the supplemental material. In the foreseeable future, we plan to make the code of our method publicly available to facilitate its adoption and further research.
\begin{figure*}
\begin{tiny}
\begin{center}
\begin{subfigure}[ht]{0.33\linewidth}
\begin{center}
\includegraphics[width=1.\linewidth]{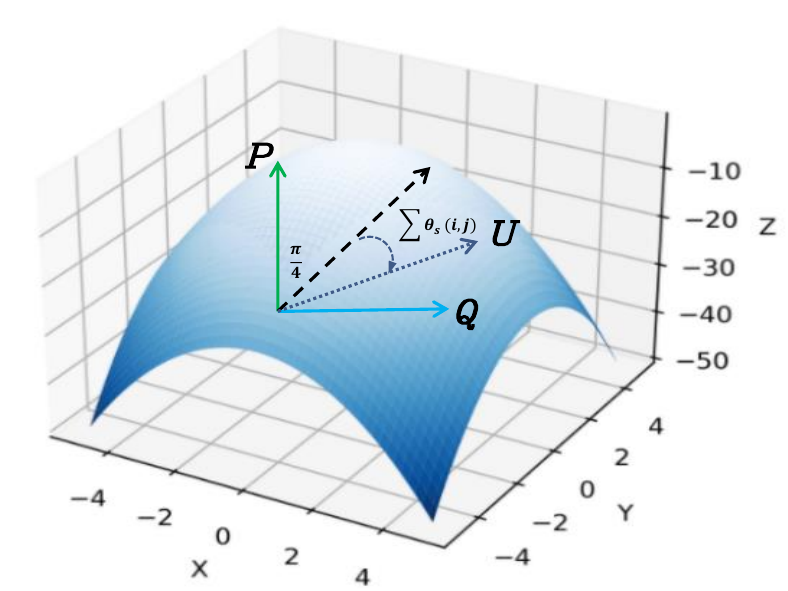}
\caption{}
\label{dm1}
\end{center}
\end{subfigure}
\begin{subfigure}[ht]{0.33\linewidth}
\begin{center}
\includegraphics[width=1.\linewidth]{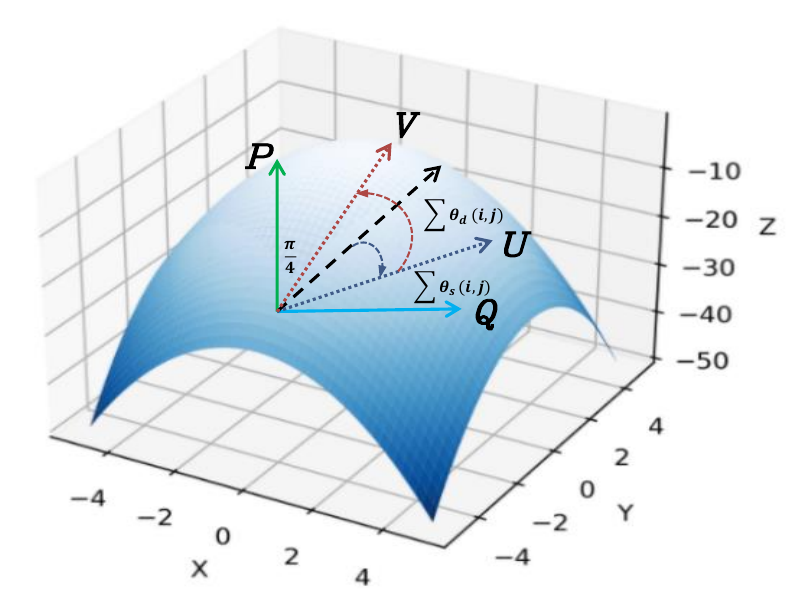} 
\caption{}
\label{dm2}
\end{center}
\end{subfigure}
\begin{subfigure}[ht]{0.33\linewidth}
\begin{center}
\includegraphics[width=1.\linewidth]{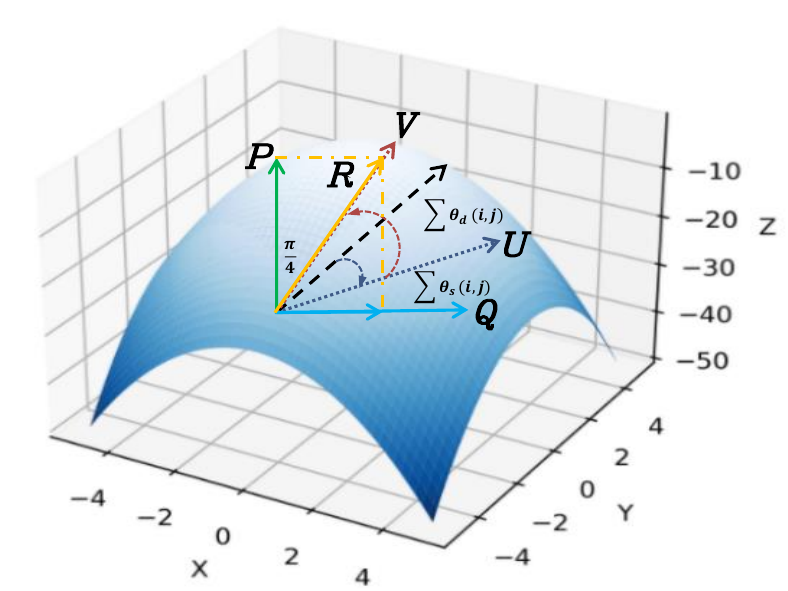} 
\caption{}
\label{dm3}
\end{center}
\end{subfigure}
\caption{The calculation process for the gradient modification direction in RWM algorithm. Firstly, we partition all categories into two groups based on their feature similarity across different tasks. The total LRR for all samples in the similar group is represented as $\sum\theta_s(i,j)$, while the total LRR for all samples in the dissimilar group is represented as $\sum\theta_d(i,j)$. As illustrated in Fig \ref{dm1} and Fig \ref{dm2}, the RWM algorithm rotates from the direction of $\frac{\pi}{4}$ to the $\mQ$ direction by $\sum\theta_s(i,j)$ and then towards the $\mP$ direction by $\sum\theta_d(i,j)$ to obtain the target direction $\mR$, as shown in Fig \ref{dm3}. During continual learning, the LRR for all samples is autonomously optimized through a self-attention mechanism.}
\label{DM}
\end{center}
% \vspace{-2em}
\end{tiny}
\end{figure*}
\section{Background}
\label{owm}
The orthogonal weight modification (OWM) algorithm is a valuable approach employed to address the issue of catastrophic forgetting in continual learning. Its primary objective is to modify the weight direction on the new task in such a way that the resulting modified direction $\mP$ becomes orthogonal to the subspace spanned by all inputs from the previous task. To construct the orthogonal projector, an iterative method resembling the Recursive Least Squares (RLS) algorithm \citep{shah1992optimal} is utilized, which needs a minimal number of previous samples.\par
We consider a feed-forward network comprising $L+1$ layers, denoted by the index $l=0,1,\cdots,L$, each employing the same activation function $g(\cdot)$. The symbol $\mathbf{\overline{x}}_l(i,j)\in \R^{s}$ represents the output of the $l$-th layer corresponding to the mean of the $i$-th batch inputs obtained from the $j$-th dataset, with $\mathbf{\overline{x}}_l(i,j)^T$ denoting the transpose matrix of $\mathbf{\overline{x}}_l(i,j)$. The computation of the modified direction $\mP$ can be expressed as follows:
\begin{equation}
\small
\begin{split}
\mP_l(i,j)&=\mP_l(i{-1},j)-\mathbf{k}_l(i,j)\mathbf{\overline{x}}_{l{-1}}(i,j)^T \mP_l(i{-1},j)
\\   \mathbf{k}_l(i,j)&=\frac{\mP_l(i{-1},j)\mathbf{\overline{x}}_{l{-1}}(i,j)}{\alpha+\mathbf{\overline{x}}_{l{-1}}(i,j)^T \mP_l(i{-1},j)\mathbf{\overline{x}}_{l{-1}}(i,j)}
\end{split}
\label{compute_P}
\end{equation}
where $\alpha$ represents a hyperparameter that decays with the number of tasks.
\section{Proposed Method}
In continual learning, some categories have a more compact feature distribution that has similar features across different tasks. For instance, in audio deepfake detection, genuine audio from different datasets has a more compact feature distribution than fake audio. To better leverage this phenomenon, we can modify the direction of the gradient based on whether or not that category shares similar features across tasks. For categories with dissimilar features across tasks, we can modify the gradient for this portion of the data in the direction orthogonal to the data plane of the old task. This ensures that learning from this portion of data in the new task does not disturb the knowledge learned from the old task. For categories with similar features across tasks, we can treat them as replay data generated from the experience-replay algorithm, which means that it is reasonable to minimize the modification of the gradient calculated from these data as much as possible.\par
\subsection{Class Regrouping}
We first consider a feed-forward network like that described in Sec. background, a deep neural network with $L+1$ layers, where $i$ is the index of the input batch, and $j$ is the index of the current task.\par
First, we compute the compactness of all categories by the average cosine distance between each two samples across all tasks, as shown in Eq \ref{comput_cosd}:
\begin{equation}
\small
    d_{r} = \frac{1}{N_r}\sum\limits_{m=1}\limits^{N_{r}}\sum\limits_{n=1}\limits^{N_{r}} cos_{dis}(x_m, x_n) \quad x_m, x_n \in class_{r}
\label{comput_cosd}
\end{equation}
where $r \in [1, R]$ is the class id, and $N_r$ represents the number of samples in $class_r$ and $cos_{dis}$ is the computing function of cosine distance. We can reasonably assume that $d_1 < d_2 < d_3 <...d_R$. Based on this assumption, we introduce a hyperparameter $r_s$, which signifies the allocation of the classes with the smallest $d_r$ values up to $r_s$ to group $\mathbb{S}$, while the remaining classes are allocated to group $\mathbb{D}$, as described in Eq \ref{group}:
\begin{equation}
\small
\begin{split}
    \mathbb{S} = \{class_1, class_2, ... class_{r_s}\}
    \\
    \mathbb{D} = \{class_{r_s}, class_{r_s + 1}, ... class_{R}\}
\end{split}
\label{group}
\end{equation}
\subsection{Self-Optimizing Direction Modification}
After splitting all classes into two groups, we calculate the modification direction $\mP$ as Eq \ref{compute_P}. The modification direction $\mP$ is a square matrix, which is orthogonal to the data plane of the old task. Then we introduce another modification direction $\mQ$, which is also a square matrix and orthogonal to $\mP$. The new direction $\mQ$ can be calculated as Eq \ref{Q_in_RWM}: 
\begin{equation}
\small
    \mQ=\mI\!-\!\mP(\mP^T \mP)^{-1} \mP^T
\label{Q_in_RWM}
\end{equation}
where the projector $\mP$, which is orthogonal to the subspace spanned by all previous inputs, can be calculated as Eq \ref{compute_P} and $\mI$ is an identity matrix.
The construction of the orthogonal projector $\mQ$ is mathematically sound \citep{haykin2002adaptive, ben2003generalized, Bengio+chapter2007}.\par
To make the model learn the adaptive modification direction automatically, a self-attention (SA) mechanism is then introduced before the classifier to obtain the attention score for each sample in a batch. The attention scores $\delta_t(i,j)$ can be calculated as Eq \ref{sin_from_att}:
\begin{equation}
\small
    [\delta_1(i,j), \delta_2(i,j), \delta_3(i,j), ... \delta_b(i,j)] = f_{SA}(h_l(i,j))
\label{sin_from_att}
\end{equation}
where $h(i, j)$ represents the hidden state of this batch before the classifier and $b$ represents the batch size. Then, all attention scores are normalized according to Eq \ref{normalize_sin}.
\begin{equation}
\small
\delta_t(i,j) = \frac{\exp{\delta_t(i,j)}}{\sum\limits_{t=1}\limits^{b} \exp{\delta_t(i,j)}}
\label{normalize_sin}
\end{equation}
We assume that each attention score $\delta_t$ can be expressed as the sine value of an angle $\theta_t$, then according to Eq \ref{theta_lower_90}, the sum of all $\theta_t$ is greater than $0$ and less than $\frac{\pi}{2}$.
\begin{equation}
\small
    0 < \sin(\sum\limits_{t=1}\limits^{b}\theta_t(i,j)) < \sum\limits_{t=1}\limits^{b}\sin\theta_t(i,j) 
\label{theta_lower_90}
\end{equation}
where $\sum\limits_{t=1}\limits^{b}\sin\theta_t(i,j) = \sin\frac{\pi}{2}$. Our algorithm adaptively adjusts the gradient modification direction for each sample based on the attention score.
The modification direction can be considered as a direction learned by the model itself, as the attention scores will be continuously optimized during model training. We name the angle $\theta_t$ as the learned rotated radians (LRR), which can be calculated according to Eq \ref{arcsin}. 
\begin{equation}
\small
    \theta_t(i,j) = \sin^{-1}(\delta_t(i,j))
\label{arcsin}
\end{equation}
For those samples belongs to a class in $\mathbb S$, such as genuine audio in audio deepfake detection, RWM first calculates the sum of their LRR in the $i$th batch of the $j$th task, denoted as $\sum \theta_s(i,j)$. For those samples belongs to a class in $\mathbb D$, such as various types of fake audio, RWM also calculates the sum of their LRR, denoted as $\sum \theta_d(i,j)$. \par
For those samples $\in \mathbb S$, as we mentioned above, we should reduce the impact of the gradient modification on them. Therefore, the gradient modification direction starts with $\frac{\pi}{4}$ and rotates towards $\mQ$ direction by $\sum\theta_s(i,j)$, obtaining a new direction $\mU$, as shown in Fig \ref{dm1}. Next, we consider those samples that have large differences in features across different tasks. The gradient modification direction starts from $\mU$ and rotates towards the $\mP$ direction by $\sum\theta_d(i,j)$, obtaining a new direction $\mV$, as shown in Fig \ref{dm2}. Here, direction $\mP$ is orthogonal to the data plane of the old task. Thus, the closer the modification direction is to $\mP$, the less interference will cause to the already learned knowledge when training on new dataset. Conversely, the closer the modification direction is to $\mQ$, which is orthogonal to the direction $\mP$, the smaller modification will be introduced during learning on a new dataset, making this process more similar to a common gradient backpropagation. After all direction modifications, we obtain the final gradient modification represented by the final LRR $\theta_f(i,j)$ as:
\begin{equation}
\small
    \theta_f(i,j) = \frac{\pi}{4} + \frac{\sum \theta_s(i,j) - \sum \theta_d(i,j)}{2}
\label{theta_m_in_RWM}
\end{equation}
where $\theta_f(i,j)$ will be optimized during the training process, so it can be viewed as a modification direction learned by the model itself. Here, we use $\frac{(...)}{2}$ to ensure that the value range of final LRR $\theta_f(i,j)$ is greater than $0$ and less than $\frac{\pi}{2}$, where the trigonometric functions are monotonous.\par
After calculating the final LRR, the final gradient modification direction $\mR$ can be easily computed based on trigonometric functions. From the Fig. \ref{dm3}, the final LRR is the angle between the direction matrix $\mP$ and $\mR$, so the direction $\mR$ can be calculated as Eq \ref{R_in_RWM}.
\begin{equation}
\small
    \mR = u(\frac{\mP}{||\mP||} + \beta\frac{\mQ}{||\mQ||}) \quad where \ u=||\mP||
\label{R_in_RWM}
\end{equation}
In Eq \ref{R_in_RWM}, $||\mP||$ and $||\mQ||$ represent the norms of $\mP$ and $\mQ$, respectively. The parameter $\beta$ is defined as the tangent value of LRR, as shown in Eq \ref{beta_in_RWM}:
\begin{equation}
\small
    \beta = \tan\theta_f
\label{beta_in_RWM}
\end{equation}
and the BP process of RWM can be written as Eq \ref{BP_RWM}: 
\begin{equation}
\small
\begin{split}
\mW_l(i,j)\!=\!\mW_l(i{-1},j)\!+\!\gamma(i,j) \Delta \mW^{BP}_l(i,j)& \qquad j\!=\!1
\\
\mW_l(i,j)\!=\!\mW_l(i{-1},j)\!+\!\gamma(i,j) \mG_l(i,j)& \qquad j\!>\!1
\\
\mG_l(i,j) = \mR_l(i,j) \Delta \mW^{BP}_l(i,j) \qquad&
\end{split}
\label{BP_RWM}
\end{equation}
where the $\Delta W_l^{BP}$ represents the gradient calculated by the standard BP algorithm. The Algorithm \ref{alg:beam-search} is also included in the paper to visually represent the implementation of our method and enhance the understanding of its structure and flow. \par
\begin{algorithm}[t]
\caption{Radian Weight Modification}
\label{alg:beam-search}
\begin{algorithmic}[1]
\begin{small}
\STATE {\bfseries Require:} Training data from different datasets, $\gamma$ (learning rate), $r_s$ (group split proportion rate).
\FOR {every class \textit{r}}
\STATE {$d_{r} = \frac{1}{N_r}\sum\limits_{m=1}\limits^{N_{r}}\sum\limits_{n=1}\limits^{N_{r}} cos_{dis}(x_m, x_n)$}
\ENDFOR
\STATE {$\mH = \mathbf{Sort}(d_1, d_2, d_3,...d_R) \qquad \vartriangleright \mH[0]\leq...\mH[R\!-\!1]$}
\STATE {$\mathbb S = \{class_r \ \mathbf{for} \ d_r \ \mathbf{in} \ \mH[:r_s]\}$}
\STATE {$\mathbb D = \{class_r \ \mathbf{for} \ d_r \ \mathbf{in} \ \mH[r_s:]\}$}
\FOR{every dataset \textit{j}}
\FOR {every batch \textit{i}}
\IF{$j =1$}
\STATE $\mW_l(i,j) = \mW_l(i{-1},j) + \gamma(i,j)\Delta \mW_l^{BP}(i,j)$
\ELSE
\STATE {$\mathbf{k}(i,j) = \dfrac{\mP_l(i{-1}) \mathbf{\overline{x}}_{l{-1}} (i,j)}{ \alpha + \mathbf{\overline{x}}_{l{-1}}(i,j)^T \mP_l(i{-1},j) \mathbf{\overline{x}}_{l{-1}} (i,j)}$}
\STATE {$\mP_l(i,j)\! =\! \mP_l(i{-1},j)\! -\! \mathbf{k}(i,j)\mathbf{\overline{x}}_{l{-1}}(i,j)^T \mP_l(i{-1},j)$} 
\STATE {$\mQ=\mI-\mP(\mP^T \mP)^{-1} \mP^T$} 
\STATE {$[\delta_1(i,j), \delta_2(i,j), \delta_3(i,j), ... \delta_b(i,j)\!= \!f_{SA}(h_l(i,j))$} 
\STATE $\sum \theta_s(i,j)=0; = \sum \theta_d(i,j) = 0$
% \vartriangleright \ \sum\limits_{t=1}\limits^{b} \delta_t(i,j)\!=\!1$}
\FOR {every sample \textit{t}}
\STATE {$\delta_t(i,j) = \frac{\exp{\delta_t(i,j)}}{\sum\limits_{t=1}\limits^{b} \exp{\delta_t(i,j)}}$}
\STATE {$\theta_t(i,j)\! =\! sin^{-1}(\delta_t(i,j)) \  \vartriangleright 0\! <\! \sum\limits_{t=1}\limits^{b}\theta_t(i,j)\! <\!\frac{\pi}{2}$}
\IF{class of $\theta_t(i,j) \in \mathbb S$}
\STATE {$\sum \theta_s(i,j) += \theta_t(i,j)$}
\ELSE
\STATE {$\sum \theta_d(i,j) += \theta_t(i,j)$}
\ENDIF
\ENDFOR
% \STATE {$\sum \theta_s(i,j) = \mathbf{Sum}(\theta_t(i,j) for )$}
\STATE {$\theta_f(i,j) = \frac{\pi}{4} + \frac{\sum \theta_s(i,j) - \sum \theta_d(i,j)}{2}$}
% \STATE {$\beta = \tan\theta_f$}
\STATE {$\mR = u(\frac{\mP}{||\mP||} + \beta\frac{\mQ}{||\mQ||}) \quad \vartriangleright u=||\mP||; \beta = \tan\theta_f$}
\STATE {$\mW_l(i,j)\!=\!\mW_l(i{-1},j)\!+\!\gamma(i,j) \mG_l(i,j)$}
\STATE {$\mG_l(i,j)= \mR_l(i,j) \Delta \mW^{BP}_l(i,j)$}
\ENDIF
\ENDFOR
\ENDFOR
\end{small}
\end{algorithmic}
\end{algorithm}
\subsection{Demonstrative Analysis}
We demonstrate the calculation formula of the direction angle $\theta_f$ for both audio deepfake detection and image recognition on the classical continual learning image recognition benchmark, CLEAR\citep{lin2021clear}. For audio deepfake detection, the compactness using pre-trained Wav2vec 2.0 \citep{2020wav2vec} feature of genuine audio $d_{genuine}$ and various types of fake audio $d_{fake}$ in the ASVspoof2019LA \citep{todisco2019asvspoof} dataset are $0.010$ and $0.062$, respectively. Obviously, the $r_s$ is 1. Under this condition, the $\theta_f$ can be written as Eq \ref{theta_m_in_RWM_FAD}.
\begin{small}
\begin{equation}
    \theta_f(i,j) = \frac{\pi}{4} + \frac{\sum \theta_1(i,j) - \sum \theta_2(i,j)}{2}
\label{theta_m_in_RWM_FAD}
\end{equation}
\end{small}
For image recognition, we calculate the in-class cosine distinance of all categories in the CLEAR dataset. The compactness $d_r$ of all categories are $\{soccer: 0.18, hockey: 0.18, bus: 0.20, baseball: 0.23, cospaly: 0.25, racing: 0.27, dress: 0.27, camera: 0.28, laptop: 0.29, 
sweater: 0.32, background: 0.40\}$. Therefore, if the hyperparameter $r_s$ is set as 4, the $\theta_f$ for this benchmark can be written as Eq \ref{theta_m_in_RWM_CLEAR}.
\begin{small}
\begin{equation}
    \theta_f(i,j) = \frac{\pi}{4} + \frac{
    \sum\limits_{c=1}\limits^{4} \sum \theta_{c}(i,j) - \sum\limits_{c=5}\limits^{11} \sum \theta_{c}(i,j)
    }
    {2}
\label{theta_m_in_RWM_CLEAR}
\end{equation}
\end{small}
In the Eq \ref{theta_m_in_RWM_FAD}, $\theta_1(i,j)$ and $\theta_2(i,j)$ represent the LRR of samples belonging to the genuine and fake categories in $i$th batch of $j$th task, respectively. In the Eq \ref{theta_m_in_RWM_CLEAR}, $\theta_c(i,j)$ ranges from $1$ to $4$, representing the LRR assigned to samples belonging to $\{soccer, hockey, bus, baseball\}$ and $\theta_c(i,j)$ ranges from $5$ to $11$, representing the LRR assigned to samples belonging to other classes.
\section{Experiments}
A series of experiments were undertaken to evaluate the efficacy of our methodology in both the audio deepfake detection and image recognition domains. In the field of audio deepfake detection, our focus was on detecting fake audio across multiple widely used datasets specifically designed for incremental synthetic algorithms audio deepfake detection. For image recognition, we employed a well-established continual learning benchmark known as CLEAR.
\subsection{Audio deepfake detection}
\subsubsection{Datasets}
We evaluate our approach on three fake audio datasets: ASVspoof2019LA ($\mathbf{S}$) \citep{todisco2019asvspoof}, ASVspoof2015 ($\mathbf{T_1}$) \citep{wu2015asvspoof}, and In-the-Wild ($\mathbf{T_2}$) \citep{muller2022does}. 
The $\mathbf{S}$ dataset includes attacks from four TTS and two VC algorithms. The bonafide audio is collected from the VCTK corpus \citep{veaux2017cstr}. The $\mathbf{T_1}$ dataset contains genuine and synthetic speech recordings from 106 speakers. The $\mathbf{T_2}$ dataset contains deep fake and genuine audio from 58 politicians and public figures collected from publicly available sources.
We constructed the training set of $\mathbf{T}_2$ by using one-third of the fake audio and an equal number of genuine audio, while the remaining audio was used as the evaluation set. The Equal Error Rate (EER) \citep{wu2015asvspoof}, which is widely used for audio deepfake detection, is applied to evaluate the performance. The detailed statistics of the datasets are presented in Table 7 in our supplementary material.
\subsubsection{Experimental Setup}:\par
\textbf{Model}: We employ the Wav2vec 2.0 model \citep{2020wav2vec} as the feature extractor, while the self-attention convolutional neural network (S-CNN) serves as the classifier. The parameters of Wav2vec 2.0 are loaded from the pre-trained model XLSR-53 \citep{conneau2020unsupervised}. The S-CNN classifier consists of three 1D-Convolution layers, one self-attention layer, and two fully connected layers in its forward process. The input dimension of the first convolution layer is 256, and all convolution layers have a hidden dimension of 80. A kernel size of 5 and a stride of 1 are applied. The fully connected layers have a hidden dimension of 80 and the output dimension of 2.\par
\textbf{Training Details}: We finetune the XLSR-53 and S-CNN using the Adam optimizer with a learning rate $\gamma$ of 0.0001 and a batch size of 2. To evaluate the performance of our proposed method for audio deepfake detection, we compared it against three widely used continual learning methods, as well as finetuning and the first continual learning method for audio deepfake detection (DFWF)\citep{ma2021continual}. In addition, we present the results of training on all datasets (Replay-All) that are considered to be the lower bound to all continual learning methods we mentioned \citep{DBLP:journals/nn/ParisiKPKW19}.
All results are (re)produced by us and averaged over 7 runs with standard deviations.
\begin{table}
\caption{The EER(\%) of our method compared with various methods. (a) and (b) are trained using the training set in order to $\mathbf{S} \rightarrow \mathbf{T_k}$ and are evaluated using the evaluation set on $\mathbf{S}$ and $\mathbf{T_k}$; (c) and (d) are trained using the training set in order to $\mathbf{S} \rightarrow \mathbf{T_1} \rightarrow \mathbf{T_2}$ and $\mathbf{S} \rightarrow \mathbf{T_2} \rightarrow \mathbf{T_1}$ and is evaluated on the evaluation set of each dataset.}
\begin{center}
\begin{subtable}[t]{0.49\linewidth}
\begin{center}
\caption{}
\label{em2a}
\resizebox{!}{1.3cm}{
\begin{tabular}{ccc}
\toprule[1.pt]
\multicolumn{1}{c}{\bf Method}  &   \multicolumn{1}{c}{$\mathbf{S}$}  &    \multicolumn{1}{c}{$\mathbf{T_1}$}
\\ 
\midrule[0.5pt]
Baseline        &   $0.258$                 &   $24.532$            \\
Replay-All      &   $0.406$                 &   $0.201$             \\
\midrule[0.5pt]
Finetune       &   $7.324$                 &   $0.510$             \\
EWC             &   $2.832$                 &   $0.570$             \\
OWM             &   $2.448$                 &   $0.540$             \\
LwF             &   $3.123$                 &   $0.343$             \\
DFWF            &   $1.849$                 &   $0.689$             \\
$\mathbf{RWM (Ours)}$ &   $\mathbf{0.438}$        &   $\mathbf{0.212}$     \\
\bottomrule[1.pt]
\end{tabular}}
\end{center}
\end{subtable}
\begin{subtable}[t]{0.49\linewidth}
\begin{center}
\caption{}
\label{em2b}
\resizebox{!}{1.3cm}{
\begin{tabular}{ccc}
\toprule[1.pt]
\multicolumn{1}{c}{\bf Method}     & \multicolumn{1}{c}{$\mathbf{S}$} & \multicolumn{1}{c}{$\mathbf{T_2}$} \\ \midrule[0.5pt]
Baseline         &             $0.258$              &              $91.473$      \\
Replay-All       &             $2.740$              &              $2.160$       \\ 
\midrule[0.5pt]
Finetune        &             $20.976$             &              $4.978$       \\
EWC              &             $8.039$              &              $5.615$       \\
OWM              &             $8.130$              &              $5.065$       \\
LwF              &             $6.453$              &              $4.998$       \\
DFWF             &             $4.324$              &              $6.275$       \\
$\mathbf{RWM (Ours)}$   & $\mathbf{3.665}$          &      $\mathbf{2.247}$      \\ 
\bottomrule[1.pt]
\end{tabular}
} 
\end{center}
\end{subtable}
\begin{subtable}[t]{0.49\linewidth}
\begin{center}
\caption{}
\label{em2c}
\resizebox{!}{1.3cm}{
\begin{tabular}{cccc}
\toprule[1.pt]
\multicolumn{1}{c}{\bf Method}  & \multicolumn{1}{c}{$\mathbf{S}$} & \multicolumn{1}{c}{$\mathbf{T_1}$}  & \multicolumn{1}{c}{$\mathbf{T_2}$} \\ 
\midrule[0.5pt]
Baseline         &             $0.258$        &         $24.532$          &              $91.473$      \\
Replay-All       &             $2.344$        &         $7.253$           &              $1.003$       \\ 
\midrule[0.5pt]
Finetune        &             $4.636$        &              $28.765$      &             $2.543$       \\
EWC              &             $8.684$        &              $12.397$      &             $3.722$       \\
OWM              &             $4.756$        &              $10.132$      &             $3.647$       \\
LwF              &             $7.505$        &              $9.547$      &              $1.540$       \\
DFWF             &             $6.211$        &              $9.672$      &              $6.478$       \\
$\mathbf{RWM (Ours)}$   & $\mathbf{2.896}$     &      $\mathbf{7.693}$     &      $\mathbf{1.161}$      \\ 
\bottomrule[1.pt]
\end{tabular}} 
\end{center}
\end{subtable}
\begin{subtable}[t]{0.49\linewidth}
\begin{center}
\caption{}
\label{em2d}
\resizebox{!}{1.3cm}{
\begin{tabular}{cccc}
\toprule[1.pt]
\multicolumn{1}{c}{\bf Method}  & \multicolumn{1}{c}{$\mathbf{S}$} & \multicolumn{1}{c}{$\mathbf{T_2}$}  & \multicolumn{1}{c}{$\mathbf{T_1}$} \\ 
\midrule[0.5pt]
Baseline         &             $0.258$        &              $91.473$     &              $24.532$           \\
Replay-All       &             $5.197$        &              $13.893$      &             $0.842$       \\ 
\midrule[0.5pt]
Finetune        &             $13.362$       &              $35.368$      &             $0.876$       \\
EWC              &             $7.343$        &              $29.516$      &             $0.933$       \\
OWM              &             $6.675$        &              $26.619$      &             $1.042$       \\
LwF              &             $10.035$       &              $32.409$      &             $0.897$       \\
DFWF             &             $6.994$        &              $24.697$       &             $1.332$       \\
$\mathbf{RWM (Ours)}$   & $\mathbf{5.616}$     &      $\mathbf{15.993}$     &      $\mathbf{0.861}$      \\ 
\bottomrule[1.pt]
\end{tabular}} 
\end{center}
\end{subtable}
\label{cm_fad}
\end{center}
\end{table}
\subsubsection{Comparison with other methods}
In this study, we evaluated the performance of our proposed method for audio deepfake detection in both two-dataset (Table \ref{em2a}, \ref{em2b}) and three-dataset (Table \ref{em2c}, \ref{em2d}) continual learning scenarios, and compared it with several other methods. The results showed that our method achieved the best detection performance compared to other methods in both scenarios, even in the presence of significant acoustic environment differences (Table \ref{em2b}). In the three-dataset continual learning scenario, our method still achieved the best performance on both old and new datasets. These results suggest that our method is effective and robust for audio deepfake detection in various continual learning scenarios with different levels of acoustic environment differences.
\subsubsection{Comparing to others with limited training samples}
\begin{table}[t]
\caption{The EER(\%) of limited samples experiments. (a), (b) and (c) are first trained using the training set of $\mathbf{S}$ and then trained on 1000, 100 and 10 samples of the training set of $\mathbf{T_2}$ respectively. All experiments are evaluated using the evaluation set on $\mathbf{S}$ and $\mathbf{T_2}$.}
\begin{center}
\label{fs_fad}
\begin{subtable}[t]{0.32\linewidth}
\begin{center}
\caption{}
\label{fs2a}
\resizebox{!}{1.cm}{
\begin{tabular}{ccc}
\toprule[1.pt]
\multicolumn{1}{c}{\bf Method}  &   \multicolumn{1}{c}{$\mathbf{S}$}  &    \multicolumn{1}{c}{$\mathbf{T_2}$}
\\ 
\midrule[0.5pt]
Baseline        &   $0.258$                 &   $91.473$            \\
Replay-All      &   $2.715$                 &   $2.162$             \\
\midrule[0.5pt]
Finetune       &   $16.437$                &   $4.999$             \\
EWC             &   $6.148$                 &   $7.576$             \\
OWM             &   $7.860$                 &   $4.364$             \\
LwF             &   $4.037$                 &   $6.391$             \\
DFWF            &   $5.129$                 &   $8.864$             \\
$\mathbf{RWM (Ours)}$ &   $\mathbf{7.670}$        &   $\mathbf{3.921}$     \\
\bottomrule[1.pt]
\end{tabular}}
\end{center}
\end{subtable}
\begin{subtable}[t]{0.32\linewidth}
\begin{center}
\caption{}
\label{fs2b}
\resizebox{!}{1.cm}{
\begin{tabular}{ccc}
\toprule[1.pt]
\multicolumn{1}{c}{\bf Method}     & \multicolumn{1}{c}{$\mathbf{S}$} & \multicolumn{1}{c}{$\mathbf{T_2}$} \\ \midrule[0.5pt]
Baseline         &             $0.258$              &              $91.473$      \\
Replay-All       &             $1.203$              &              $4.198$       \\ 
\midrule[0.5pt]
Finetune        &             $16.058$             &              $6.503$       \\
EWC              &             $7.666$              &              $8.977$       \\
OWM              &             $8.229$              &              $8.177$       \\
LwF              &             $5.750$              &              $5.950$       \\
DFWF             &             $4.246$              &              $9.879$       \\
$\mathbf{RWM (Ours)}$   & $\mathbf{1.507}$           &      $\mathbf{5.305}$      \\ 
\bottomrule[1.pt]
\end{tabular}
} 
\end{center}
\end{subtable}
\begin{subtable}[t]{0.32\linewidth}
\begin{center}
\caption{}
\label{fs2c}
\resizebox{!}{1.cm}{
\begin{tabular}{ccc}
\toprule[1.pt]
\multicolumn{1}{c}{\bf Method}  & \multicolumn{1}{c}{$\mathbf{S}$}  & \multicolumn{1}{c}{$\mathbf{T_2}$} \\ 
\midrule[0.5pt]
Baseline         &             $0.258$          &              $91.473$      \\
Replay-All       &             $0.897$          &              $15.326$       \\ 
\midrule[0.5pt]
Finetune        &             $8.223$          &             $19.385$       \\
EWC              &             $7.301$          &             $18.599$       \\
OWM              &             $7.021$          &             $19.684$       \\
LwF              &             $8.019$          &             $19.673$       \\
DFWF             &             $6.894$          &             $19.992$       \\
$\mathbf{RWM (Ours)}$   & $\mathbf{2.463}$       &     $\mathbf{17.252}$      \\ 
\bottomrule[1.pt]
\end{tabular}} 
\end{center}
\end{subtable}
\end{center}
\end{table}
To verify the sensitivity of our method to the amount of training data for new tasks in continual learning, we conducted experiments with different numbers of training data for new tasks and compared our method with others, as shown in Table \ref{fs_fad}. The results show that our method performs better than other continual learning methods on new tasks with less training data, and generally has better performance in mitigating forgetting compared to other methods. However, when the number of training data decreases from 100 (Table \ref{fs2b}) to 10 (Table \ref{fs2c}), the ability of our method to mitigate forgetting decreases. This is because our method requires data to allow the model to learn the appropriate gradient modification direction. If the amount of training data is too small, the model may not learn the optimal modification direction, resulting in poorer performance on the old dataset.
\subsubsection{Ablation studies for our method}
\begin{table}
\caption{The EER(\%) on evaluation sets of the ablation studies. (a) and (b) are trained using the training set in order to $\mathbf{S} \rightarrow \mathbf{T_k}$ and are evaluated using the evaluation set on $\mathbf{S}$ and $\mathbf{T_k}$; (c) and (d) are trained using the training set in order to $\mathbf{S} \rightarrow \mathbf{T_1} \rightarrow \mathbf{T_2}$ and $\mathbf{S} \rightarrow \mathbf{T_2} \rightarrow \mathbf{T_1}$ and is evaluated using evaluation sets.}
\begin{center}
\begin{subtable}[t]{0.49\linewidth}
\begin{center}
\caption{}
\label{ab2a}
\resizebox{!}{0.8cm}{
\begin{tabular}{ccc}
\toprule[1.pt]
\multicolumn{1}{c}{\bf Method}  &   \multicolumn{1}{c}{$\mathbf{S}$}  &    \multicolumn{1}{c}{$\mathbf{T_1}$}
\\ 
\midrule[0.5pt]
Baseline             &   $0.258$                 &   $24.532$            \\
\midrule[0.5pt]
$\mathbf{RWM (Ours)}$ &   $\mathbf{0.438}$   &   $\mathbf{0.212}$     \\
--LRR                &   $2.448$            &   $0.540$             \\
--WM                 &   $7.324$            &   $0.510$             \\
\bottomrule[1.pt]
\end{tabular}}
\end{center}
\end{subtable}
\begin{subtable}[t]{0.49\linewidth}
\begin{center}
\caption{}
\label{ab2b}
\resizebox{!}{0.8cm}{
\begin{tabular}{ccc}
\toprule[1.pt]
\multicolumn{1}{c}{\bf Method}     & \multicolumn{1}{c}{$\mathbf{S}$} & \multicolumn{1}{c}{$\mathbf{T_2}$} \\ \midrule[0.5pt]
Baseline         &             $0.258$              &              $91.473$      \\ 
\midrule[0.5pt]
$\mathbf{RWM (Ours)}$   & $\mathbf{3.665}$          &      $\mathbf{2.247}$      \\
--LRR                  &             $8.130$              &              $5.065$       \\
--WM                   &             $20.976$             &              $4.978$       \\
\bottomrule[1.pt]
\end{tabular}
} 
\end{center}
\end{subtable}
\begin{subtable}[t]{0.49\linewidth}
\begin{center}
\caption{}
\label{ab2c}
\resizebox{!}{0.8cm}{
\begin{tabular}{cccc}
\toprule[1.pt]
\multicolumn{1}{c}{\bf Method}  & \multicolumn{1}{c}{$\mathbf{S}$} & \multicolumn{1}{c}{$\mathbf{T_1}$}  & \multicolumn{1}{c}{$\mathbf{T_2}$} \\ 
\midrule[0.5pt]
Baseline         &             $0.258$        &         $24.532$          &              $91.473$      \\ 
\midrule[0.5pt]
$\mathbf{RWM (Ours)}$   & $\mathbf{2.896}$     &      $\mathbf{7.693}$     &      $\mathbf{1.161}$      \\
--LRR              &             $4.756$        &              $10.132$      &             $3.647$       \\
--WM        &             $4.636$        &              $28.765$      &             $2.543$       \\
\bottomrule[1.pt]
\end{tabular}} 
\end{center}
\end{subtable}
\begin{subtable}[t]{0.49\linewidth}
\begin{center}
\caption{}
\label{ab2d}
\resizebox{!}{0.8cm}{
\begin{tabular}{cccc}
\toprule[1.pt]
\multicolumn{1}{c}{\bf Method}  & \multicolumn{1}{c}{$\mathbf{S}$} & \multicolumn{1}{c}{$\mathbf{T_2}$}  & \multicolumn{1}{c}{$\mathbf{T_1}$} \\ 
\midrule[0.5pt]
Baseline         &             $0.258$        &              $91.473$     &              $24.532$           \\
\midrule[0.5pt]
$\mathbf{RWM (Ours)}$   & $\mathbf{5.616}$     &      $\mathbf{15.993}$     &      $\mathbf{0.861}$      \\
--LRR              &             $6.675$        &              $26.619$      &             $1.042$       \\
--WM        &             $13.362$       &              $35.368$      &             $0.876$       \\
\bottomrule[1.pt]
\end{tabular}} 
\end{center}
\end{subtable}
\label{ab_fad}
\end{center}
\end{table}
% In order to evaluate the effectiveness of our method for audio deepfake detection, 
We also conducted ablation experiments similar to image recognition, as shown in Table \ref{ab_fad}. From the results, we can observe that continual learning on the new dataset without the LRR and WM can cause the model to disrupt previously learned knowledge, resulting in an increase in error rate on the old dataset, particularly evident in different acoustic environments of the new and old datasets as shown in Table \ref{ab2b}, \ref{ab2c} and \ref{ab2d}. The results demonstrate that the gradient direction modification mechanism has a positive effect on overcoming forgetting in most experimental settings. However, this mechanism also reduces the learning performance on new tasks. Additionally, we observed that the self-learning mechanism of gradient-modified radian introduced in our method has a positive impact on the performance of both overcoming forgetting and acquiring new knowledge in all experimental settings. Furthermore, it significantly alleviates the recognition performance loss caused by the introduction of the gradient direction modification on new tasks.
\subsection{Image Recognition}
\begin{figure}[t]
\begin{center}
\includegraphics[width=0.8\linewidth]{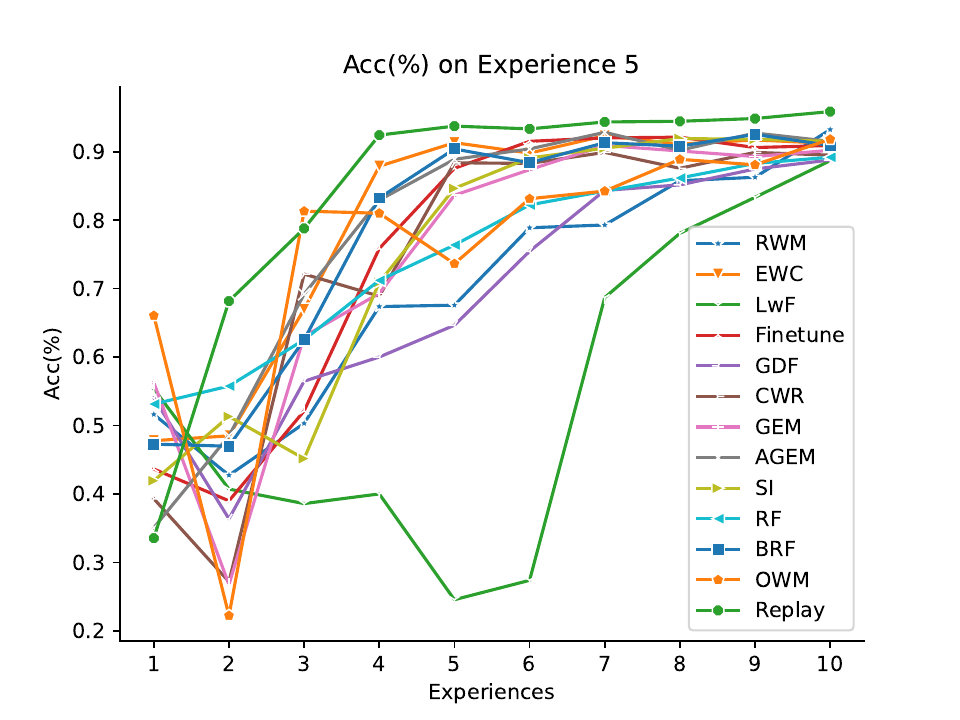}
\caption{The performance of different continual learning methods after training 5 experiences of CLEAR. All methods are trained using the training sets of $\mathbf{Exp}_1$ to $\mathbf{Exp}_{5}$ in sequence. The accuracy of all methods in all experiences has been added to the supplemental material.} 
\label{acc_matrix}
\end{center}
\end{figure}
\subsubsection{Dataset}
We use the CLEAR benchmark to evaluate the performance of our method for image recognition. CLEAR is a classical continual learning benchmark that is based on the natural temporal evolution of visual concepts of Internet images. Task-based sequential learning is adopted with a sequence of 10-way classification tasks by splitting the temporal stream into 10 buckets, each consisting of a labeled subset for training and evaluation. A small labeled subset ($\mathbf{Exp}_1$, $\mathbf{Exp}_2$, $\mathbf{Exp}_3$, ... $\mathbf{Exp}_{10}$) consisting of 11 temporally dynamic categories with 300 labeled images per category, which includes illustrative categories such as computer, cosplay, etc., as well as a background category. In continual learning, only the current task data is available at each timestamp, except for the replay-based algorithm. The train and evaluation datasets of each labeled subset are generated by using the classic 70/30\% train-test split as Table 6 in our supplementary material.
\begin{table*}
% \vspace{-1em}
\caption{The accuracy(\%) of the model after training on all CLEAR experiences. All results are (re)produced by us and averaged over 7 runs with standard deviations. The full details of all methods have been described in supplementary material.}
\label{Compared_CLEAR}
\begin{center}
\resizebox{!}{3.0cm}{
\begin{tabular}{ccccccccccc}
\toprule[1.pt]
\multirow{2.5}{*}{\bf Continual Learning Methods} & 
\multicolumn{10}{c}{\bf Accuracy on each experience}         \\
\cmidrule[0.5pt]{2-11}
~ & $\mathbf{Exp_1}$ & $\mathbf{Exp_2}$ & $\mathbf{Exp_3}$ & $\mathbf{Exp_4}$ & $\mathbf{Exp_5}$ & $\mathbf{Exp_6}$ & $\mathbf{Exp_7}$ & $\mathbf{Exp_8}$ & $\mathbf{Exp_9}$ & $\mathbf{Exp_{10}}$\\ 
\midrule[0.5pt]
Replay-All & $\mathbf{94.85}$ & $\mathbf{94.65}$ & $\mathbf{94.75}$ & $\mathbf{94.65}$ & $\mathbf{95.86}$ & $\mathbf{95.35}$ & $\mathbf{95.15}$ & $\mathbf{94.65}$ & $\mathbf{95.76}$ & $\mathbf{96.16}$\\
\midrule[0.5pt]
Finetune & $87.68$ & $90.00$ & $91.11$ & $91.82$ & $90.40$ & $89.90$ & $90.30$ & $90.61$ & $90.61$ & $93.33$\\
EWC & $84.04$ & $84.95$ & $85.86$ & $87.07$ & $85.66$ & $85.56$ & $86.97$ & $86.16$ & $85.76$ & $87.78$\\
LwF & $88.59$ & $88.89$ & $87.27$ & $90.51$ & $87.68$ & $87.78$ & $87.47$ & $87.47$ & $88.79$ & $88.48$\\
GDF & $91.11$ & $91.62$ & $88.38$ & $91.01$ & $88.79$ & $89.19$ & $90.20$ & $87.68$ & $90.10$ & $90.30$\\
CWR & $90.71$ & $91.72$ & $90.71$ & $91.52$ & $89.49$ & $90.91$ & $91.62$ & $90.71$ & $91.82$ & $93.74$\\
GEM & $88.38$ & $89.70$ & $90.81$ & $91.41$ & $90.20$ & $89.29$ & $90.91$ & $89.60$ & $90.71$ & $93.03$\\
AGEM & $92.32$ & $91.41$ & $92.02$ & $93.43$ & $91.52$ & $92.32$ & $92.22$ & $91.52$ & $92.83$ & $94.75$\\
SI & $91.31$ & $92.02$ & $91.41$ & $93.74$ & $91.52$ & $91.72$ & $92.63$ & $91.11$ & $92.22$ & $95.05$\\
BRF & $89.29$ & $88.99$ & $88.18$ & $88.48$ & $89.19$ & $89.19$ & $90.10$ & $88.38$ & $89.19$ & $90.00$\\
RF & $88.38$ & $90.30$ & $90.00$ & $91.62$ & $90.91$ & $90.61$ & $91.41$ & $91.41$ & $91.11$ & $93.33$\\
OWM & $91.62$ & $92.12$ & $91.82$ & $93.64$ & $91.72$ & $92.42$ & $92.22$ & $92.32$ & $92.42$ & $95.05$\\
\midrule[0.5pt]
% RAWM & $92.12$ & $92.53$ & $91.41$ & $93.74$ & $91.82$ & $92.42$ & $92.53$ & $92.22$ & $92.53$ & $\mathbf{95.25}$\\
RWM (Ours, $r_s\!=\!3$) & $92.15$ & $\mathbf{94.12}$ & $92.34$ & $93.48$ & $\mathbf{94.91}$ & $93.01$ & $92.78$ & $93.52$ & $93.76$ & $92.70$ \\
\textbf{RWM (Ours, $r_s\!=\!4$)} & $\mathbf{93.64}$ & $93.64$ & $\mathbf{92.53}$ & $\mathbf{93.84}$ & $93.23$ & $\mathbf{93.13}$ & $\mathbf{92.93}$ & $\mathbf{93.03}$ & $\mathbf{94.14}$ & $\mathbf{95.25}$\\
RWM (Ours, $r_s\!=\!5$) & $93.35$ & $92.95$ & $93.01$ & $92.75$ & $92.62$ & $92.45$ & $93.17$ & $92.64$ & $93.96$ & $93.68$\\
% Model-5  &   $0.259$   &   $25.698$   &  $44.741$   &   $91.824$  \\
% Model-6  &   $0.262$   &   $27.872$   &  $49.726$   &   $92.113$  \\
\bottomrule[1.pt]
\end{tabular}
}
\end{center}
\end{table*}
\begin{table*}[t]
% \vspace{-1em}
\caption{The ablation study of our method. All results are the accuracy(\%) on the CLEAR experiences. }
\label{ABL_CLEAR}
\begin{center}
\resizebox{!}{1.0cm}{
\begin{tabular}{ccccccccccc}
\toprule[1.pt]
\multirow{2.5}{*}{\bf Ablation study} & 
\multicolumn{10}{c}{\bf Accuracy on each experience}         \\
\cmidrule[0.5pt]{2-11}
~ & $\mathbf{Exp_1}$ & $\mathbf{Exp_2}$ & $\mathbf{Exp_3}$ & $\mathbf{Exp_4}$ & $\mathbf{Exp_5}$ & $\mathbf{Exp_6}$ & $\mathbf{Exp_7}$ & $\mathbf{Exp_8}$ & $\mathbf{Exp_9}$ & $\mathbf{Exp_{10}}$\\ 
\midrule[0.5pt]
% Replay & $\mathbf{94.85}$ & $\mathbf{94.65}$ & $\mathbf{94.75}$ & $\mathbf{94.65}$ & $\mathbf{95.86}$ & $\mathbf{95.35}$ & $\mathbf{95.15}$ & $\mathbf{94.65}$ & $\mathbf{95.76}$ & $\mathbf{96.16}$\\
% \midrule[0.5pt]
\textbf{RWM (Ours)} & $\mathbf{93.64}$ & $\mathbf{93.64}$ & $\mathbf{92.53}$ & $\mathbf{93.84}$ & $\mathbf{93.23}$ & $\mathbf{93.13}$ & $\mathbf{92.93}$ & $\mathbf{93.03}$ & $\mathbf{94.14}$ & $\mathbf{95.25}$\\
--LRR & $91.62$ & $92.12$ & $91.82$ & $93.64$ & $91.72$ & $92.42$ & $92.22$ & $92.32$ & $92.42$ & $95.05$\\
--WM & $87.68$ & $90.00$ & $91.11$ & $91.82$ & $90.40$ & $89.90$ & $90.30$ & $90.61$ & $90.61$ & $93.33$\\
\bottomrule[1.pt]
\end{tabular}}
\end{center}
\end{table*}
\subsubsection{Experimental Setup}
In the image recognition experiment, all continual learning methods are conducted using an upstream-downstream framework. The upstream component utilized the default pre-trained ResNet 50 \citep{he2016deep} of torchvision as a feature extractor, which will be frozen during the continual learning process, producing 2048-dimensional features. The downstream classifier was a linear layer with input and output dimensions of 2048 and 11, respectively. The experiment used a batch size of 512 and an initial learning rate of 1, which decayed by a factor of 0.1 after 60 epochs. We employed the SGD optimizer with a momentum of 0.9. The $\alpha$ in Eq \ref{compute_P} is 0.1 and the norm in Eq \ref{R_in_RWM} is $\mL^2$ norm. The details of all continual learning methods have been described in supplementary material.\par
\subsubsection{Comparison with other methods}
In this experiment, we compared our method RWM with several other continual learning methods with three regrouping hypermeters $r_s$ of Eq. \ref{comput_cosd} in Tabel. \ref{Compared_CLEAR}. As shown in this table, the performance of our method was second only to Replay-All after training on all experiences, which is considered the upper bound of continual learning performance. However, from Fig \ref{acc_matrix}, it can be seen that our method had lower accuracy than most of the other continual learning methods before the experience 8. This is because our method requires the model to learn the direction of gradient modification by itself. Therefore, the model not only needs to learn to discriminate input data, but also needs to learn the modified direction for different sample data on different tasks, which is the major limitation of RWM. The results also demonstrate the influence of varying $r_s$ on the outcomes. Thus, determining the optimal $r_s$ stands as a crucial avenue for our forthcoming research works.
\subsubsection{Ablation study for our method}
we also conducted an ablation study to evaluate the efficacy of our proposed method. The findings, presented in Table \ref{ABL_CLEAR}, demonstrate that both the self-learned gradient-modified radian and the gradient direction modification introduced in our method positively impact recognition performance.
Notably, our observations reveal that, in most cases, the gain in recognition performance resulting from the gradient direction modification exceeds that achieved through the self-learning of the modification radian mechanism. This observation suggests the potential for exploring more refined strategies for learning modification radian in future endeavors.
The comprehensive outcomes of our ablation study furnish compelling evidence substantiating the effectiveness of our proposed method. Moreover, they underscore the significance of the self-learned gradient-modified radian and the gradient direction modification in attaining superior recognition performance in the context of continual learning.
\section{Conclusion}
%\color{blue}
This paper proposes an effective continual learning algorithm, Radian Weight Modification (RWM), designed to enhance the adaptability and resilience of audio deepfake detection models in the face of emerging and diverse attack types. The core principle of RWM revolves around the insightful categorization of classes into two distinct groups based on feature distribution similarities. This strategic partitioning enables the algorithm to dynamically adjust gradient modification directions, effectively balancing the acquisition of new knowledge and the preservation of previously learned information across tasks. The experimental results showcased the remarkable effectiveness of RWM in comparison to mainstream continual learning methods for audio deepfake detection, signifying its robustness in addressing the challenges posed by new deepfake attack types. In addition, RWM also demonstrates successful extension to diverse machine learning domains, notably image recognition. 
Looking forward, it would be interesting to investigate how our algorithm can be extended to address other related problems in machine learning, such as domain adaptation, transfer learning, and multi-task learning \citep{threat2, threat1}.
\section{Acknowledgments}
This work is supported by the Scientific and Technological Innovation Important Plan of China (No. 2021ZD0201502), the National Natural Science Foundation of China (NSFC) (No. 62322120, No. 62306316, No.61831022, No.U21B2010, No.62101553, No.61971419, No.62006223, No. 62206278).
\bibliography{aaai24}

\end{document}

%% file: math_commands.tex
\usepackage{amsmath,amsfonts,bm}

\def\eqref#1{equation~\ref{#1}}
\def\1{\bm{1}}

\def\mG{{\bm{G}}}
\def\mH{{\bm{H}}}
\def\mI{{\bm{I}}}

\def\mL{{\bm{L}}}

\def\mP{{\bm{P}}}
\def\mQ{{\bm{Q}}}
\def\mR{{\bm{R}}}

\def\mU{{\bm{U}}}
\def\mV{{\bm{V}}}
\def\mW{{\bm{W}}}

\DeclareMathAlphabet{\mathsfit}{\encodingdefault}{\sfdefault}{m}{sl}
\SetMathAlphabet{\mathsfit}{bold}{\encodingdefault}{\sfdefault}{bx}{n}

\newcommand{\R}{\mathbb{R}}